\documentstyle[12pt]{article}

\newcommand{\be}{\begin{equation}}
\newcommand{\e}{\end{equation}}
\newcommand{\la}{\lambda}
\newcommand{\vi}{\varphi}
\newcommand{\ww}{{\cal W}}
\newcommand{\is}{\int\!\!\!\!\!\!\!\mbox{$\sum$}}
\newcommand{\mif}[2]{\mbox{$\frac{#1}{#2}$}}

\begin{document}
\setlength{\baselineskip}{20pt}

\thispagestyle{empty}
{\parindent0em DESY 94-025 \\
February 1994}
\begin{center}
\vspace{3cm}
{\Large\bf Finite Temperature Effective Potential \\
to Order $g^4,\la^2$ and \\
the Electroweak Phase Transition\\}
\vspace{2cm}
Zolt\'an Fodor
\footnote{On leave from Institute for Theoretical Physics,
E\"otv\"os University, Budapest, Hungary}
 and Arthur Hebecker\\ \vspace{.3cm}
{\small\it Deutsches Elektronen-Synchrotron DESY, Hamburg D-22603, Germany\\}
\vspace{3cm}
{\bf Abstract}
\end{center}
\begin{quote}
The standard model effective potential is calculated at finite
temperature to order $g^4,\la^2$ and a complete zero temperature
renormalization is performed. In comparison with lower order calculations the
strength of the first order phase transition has increased dramatically.
This effect can be traced back
to infrared contributions from typical non-Abelian diagrams
and to the infrared behaviour of the scalar sector close to the
critical temperature. Several quantities, e.g. surface tension, latent heat and
field expectation value are analyzed for an SU(2)-Higgs model and for the full
standard model in detail. An explicit formula
enabling further analytic or numerical study is presented.
\end{quote}
\newpage

\setlength{\baselineskip}{19pt}
\section{Introduction}
Recently it has been clarified that the electroweak phase transition
plays an important role for the observed  baryon asymmetry
of the universe \cite{KRS} (for a recent review see \cite{CKN}).

Several approaches have been used to determine the details of the
electroweak phase transition. Important results have been obtained by use of
3-dimensional effective theory \cite{PATKOS},\linebreak
$\epsilon$-expansion \cite{MR,AY} and average action \cite{RW}.
There is also a growing interest in
lattice simulations of the phase transition \cite{LATTICE}.

Perturbative calculations of the finite temperature effective potential
of the standard model have been carried
out using the one loop ring summation \cite{Ca,BFHW} to order
$g^3,\lambda^{3/2}$\linebreak ($g$ denotes the gauge-coupling and the
top-Yukawa coupling). Two-loop summation has been
done to order $g^4,\lambda$ in \cite{AE}, in which scalar masses
have been neglected with respect to gauge-boson masses, and by use of
another  approximation in  \cite{BD}.

However, there is a need to extend the work of Arnold and Espinosa \cite{AE} to
a complete $g^4,\la^2$-calculation. The analysis of the Abelian Higgs model has
shown, that higher order $\la$-corrections can change the $g^4,\la$-result for
the potential and surface tension significantly \cite{He}. Despite the general
scepticism concerning the predictive power of the perturbative results in the
$\Phi^4$-theory, note that calculating far enough in the resummed loop
expansion the obtained effective potential suggests the correct second order
phase transition. For the $Z_2$-symmetric $\Phi^4$-model one should go to the
order $\la^2$, while for the O(4)-symmetric $\Phi^4$-model even a
$\la^{3/2}$-calculation is sufficient.

The calculation of the $g^4,\la^2$-potential for the standard model and its
detailed analysis are the main goals of the present paper.
A systematic expansion in coupling constants is performed taking into account
the effects of infrared divergences \cite{BFHW} and keeping the full
dependence on the Higgs field $\vi$, its zero temperature vacuum expectation
value $v$ and the temperature T. The effect of the
higher order $\la$-corrections is found to be
important for realistic Higgs masses.

The principal method of the calculation, based on the Dyson-Schwinger equation
for the derivative $\partial V/\partial\vi$, is explained in sect. 1 for some
general theory containing all the
important features of the standard model. Essentially a
summation of tadpole diagrams is performed \cite{WKL}. The application
to the standard model is carried out and the
renormalization at zero temperature is presented.

In sect. 2 the pure SU(2)-Higgs model is analysed in detail. Using the
surface tension and other physical quantities a comparison of the
different order calculations is performed.
\setlength{\baselineskip}{20pt}
In contrast to the Abelian case,
here the $g^4,\la$-potential suggests a stronger first order phase
transition than the $g^3,\la^{3/2}$-potential. Improving the calculation from
order $g^4,\la$ to $g^4,\la^2$ a stronger first order phase transition is
obtained for both the Abelian and the non-Abelian case.
Of course, this effect questions the reliability of
the perturbative approach. The increase of the surface tension is traced back
to the infrared features of typical non-Abelian diagrams. The observed
numerical importance of the $\la$-corrections has its roots in the infrared
region as well.

The complete standard model results are discussed in sect. 3.
The large top quark mass
leads to a decrease of the surface tension. Nevertheless,
the qualitative picture is the same as for the pure SU(2) case.

After some conclusions in sect. 4 the complete analytic result for the standard
model is presented in the appendix.

\section{Calculation of the effective potential at finite temperature}
\subsection{General idea}

The effective potential $V$ is calculated using Dyson-Schwinger equations,
as described for the Abelian Higgs model in \cite{He}.
A similar way of summing the
different contributions to $V$ for the $\Phi^4$ theory
has been considered in \cite{BBH}.

Consider a general Lagrangian with interaction terms generating 3- and
4-vertices proportional to $g^2$ and 3-vertices proportional to $g\,k_\mu$. A
generic coupling constant $g$ is used as an expansion parameter and $k_\mu$
is a momentum variable. Note that this structure is suggested by the standard
model
Lagrangian where the square root of the scalar coupling $\sqrt\la$, the
Yukawa coupling $g_Y$, the electroweak gauge couplings $g_1,g_2$ and the
strong gauge coupling $g_s$ play the role of the generic coupling $g$.
Here we give all contributions to the finite temperature effective potential up
to order $g^4$. All calculations are carried out in the imaginary time
formalism.

Using the well known technique of Dyson-Schwinger equations the following
relation can be obtained for the effective potential\\
\be\hspace*{9cm}.\label{0}\e
Here the internal lines represent all particles of the theory and $\vi$ is the
``shift'' of the Lagrangian in the scalar sector. The two different sorts of
blobs are full propagator and full 3-vertex respectively. The first term gives
    \be A=\mbox{tr}\,\ww(\vi)\is\frac{dk}{k^2+m_{tree}^2+\Pi(k)}\, .\label{1}\e
In general, mass, self energy and vertex $\ww$ are matrices.
``tr'' denotes the sum over the suppressed indices.
For vector particles the self energies are different for the longitudinal
and transverse part. They can be calculated using the corresponding projection
operators (see \cite{K,BHW}). The $\vi$-dependence of the propagators is
obvious.

The Dyson-Schwinger equation for $\Pi(k)$, to the order needed in this
calculation, reads\\
\be\hspace*{9cm}.\e
In the following the indices 2 and 3 denote the contributions of order $g^2$
and $g^3$ respectively. The tadpole part of the self energy can be written as
    \be \Pi_a(k)=\Pi_{a2}+\Pi_{a2}(k)+\Pi_{a3}+\cdots\qquad\mbox{with}\quad
    \Pi_{a2}(0)=0\, .\label{5}\e
In the standard model the only nonvanishing $\Pi_{a2}(k)$-contribution is the
longitudinal self energy of a non-Abelian gauge boson. It is introduced by
the corresponding projection operator when applied to the four vector vertex.
The momentum dependence of the third order term disappears if $k_0=0$, thus
in the order we are calculating it can be neglected.

The other part of the self energy, $\Pi_b(k)$, has no contribution of order
$g^2$ for scalars. Nevertheless, for gauge particles those terms
do appear. The leading order momentum independent part of $\Pi_b(k)$ will be
called $\Pi_{b2}$.

Using these definitions and introducing the corrected mass term $m^2$,
    \be m^2=m_{tree}^2+\Pi_{a2}+\Pi_{b2}\quad, \e
equation (\ref{1}) can be written as
    \begin{eqnarray}\!\!\!\!\!A&\!\!\!=&\!\!\!\mbox{tr}\,\ww(\vi)\is\frac{dk}
    {k^2+m^2+\Pi_{a2}(k)+\Pi_{a3}+\Pi_b(k)-\Pi_{b2}}\nonumber\\
\!\!\!\!\!&\!\!\!=&\!\!\!\mbox{tr}\,\ww(\vi')\is dk\Bigg(\frac{1}{k^2+m^2+
    \Pi_{a2}(k)}-\frac{1}{k^2+m^2}\Pi_{a3}\frac{1}{k^2+m^2}\nonumber\\
\!\!\!\!\!&\!\!\!&\!\!\!  +\frac{1}{k^2+m^2}\Pi_{b2}\frac{1}{k^2+m^2}
    -\frac{1}{k^2+m^2}\Pi_b(k)\frac{1}{k^2+m^2}\Bigg)\,
    .\label{6}\end{eqnarray}
Here the second equality is obtained by expanding the integrand in $g$. This is
best seen by considering the $k_0=0$ and $k_0\neq 0$ parts separately.

Observe that in term $B$ of (\ref{0}) the vertex need not be corrected to
obtain the full $g^4$-result. Inspection of the last term of
(\ref{6}) and term $B$ of (\ref{0}) shows that their sum is equal to the
derivative $-\partial V_\ominus/
\partial\vi$, where $-V_\ominus$ represents the sum of all two-loop diagrams of
the type shown in fig. 1.a (setting sun diagrams).

With the definitions
    \begin{eqnarray}\!\!\!\!\!V_R&\!\!\!=&\!\!\!-\int^\vi d\vi'
    \mbox{tr}\,\ww(\vi')\is dk\left(\frac{1}{k^2+m^2+
    \Pi_{a2}(k)}+\frac{1}{k^2+m^2}\Pi_{b2}\frac{1}{k^2+m^2}
    \right)\, ,\label{3}\\
\!\!\!\!\!V_z&\!\!\!=&\!\!\!\quad\int^\vi d\vi'\mbox{tr}\,\ww(\vi')\is dk
    \frac{1}{k^2+m^2}\Pi_{a3}\frac{1}{k^2+m^2}\, ,\label{7}\end{eqnarray}
the potential can be given in the form
    \be V=V_{tree}\!+V_\ominus+V_z+V_R\, .\e
Note that $V_z$ can be identified as the sum of all terms bilinear in masses
coming from two-loop diagrams of the type shown in fig. 1.b.

Denoting by $V_3$ the sum of the tree level potential and the $g^3$-order part
of $V_R$ and calling $V_4$ the fourth order corrections of $V_R$,
    \be V_3+V_4=V_{tree}+V_R, \e
the following final formula is obtained :
    \be V=V_3+V_4+V_\ominus+V_z. \label{8}\e

It is worthwhile to mention the differences between the method given here and
the one presented in \cite{AE}. One advantage of our approach is the absence of
thermal counterterms. The other one is the fact that no different treatment
of zero and nonzero Matsubara frequency modes is needed. Nevertheless,
performing the rather tedious calculation using both methods the above
mentioned advantages turned out to be marginal.

\subsection{Standard model calculation}
To fix our notation the essential parts of the Lagrangian are given
    \be \cal{L}=\cal{L}_{Higgs}+\cal{L}_{gauge}+\cal{L}_{fermion}
    +\cal{L}_{Yukawa}. \e
Defining the covariant derivative as
    \be D_\mu=\partial_\mu+ig_1\frac{Y}{2}B_\mu+ig_2\frac{\tau^a}{2}W^a_\mu \e
the fermionic and gauge parts are unambiguous. The Higgs contribution reads
    \be \cal{L}_{Higgs}=-|D_\mu\Phi|^2+\nu|\Phi|^2-\la|\Phi|^4\quad,\quad
    \mbox{where}\quad\Phi=\frac{1}{\sqrt{2}}
    \left(\begin{array}{c}\vi_3+i\vi_4\\ \vi+\vi_1+i\vi_2\end{array}\right)\e
denotes the Higgs doublet.
All fermions except the top quark are considered to be massless. The resulting
Yukawa Lagrangian reads
    \be \cal{L}_{Yukawa}=-g_Y\bar{q}_L\tilde{\Phi}t_R\, ,\qquad q_L=\left(
    \begin{array}{c}t_L\\b_L\end{array}\right),\qquad\tilde{\Phi}=i\tau_2
    \Phi^*.\e

The calculation is performed in Landau gauge. To define the potential to the
order $g^4,\la^2$ the formal power counting rule
    \be g_1\sim g_2\sim g_Y\sim\la^{1/2} \e
is used. We assume $\nu=\la v^2$ to be of order $\la$, where $v$ is the zero
temperature vacuum expectation value of the scalar field.
This expansion in the couplings seems to
be natural, since it corresponds to the general
structure of the theory, described in the previous subsection. All masses
are treated symmetrically as terms of order $g$. To the given order the full
dependence on temperature, order parameter $\vi$ and $v$ is kept.

This clarifies the way, how the general considerations of the previous
subsection have to be applied to the standard model. The different
contributions to $V_\ominus$ are shown in\linebreak
fig. 2. As usual, solid,
dashed and wavy lines represent fermion, scalar and vector propagators
respectively. In order to make the explicit comparison with the results of
\cite{AE} easier we follow the labeling of the setting sun diagrams given
there
    \be V_\ominus=V_a+V_b+V_i+V_j+V_m+V_p .\label{9}\e
We have included ghost contributions in $V_m$. $V_p$ is the scalar setting sun
contribution not considered in the standard model calculation of \cite{AE}.

The calculation needed for the temperature dependent masses to order
$g^3,\la^{3/2}$ is similar to that performed in \cite{BFHW,BHW}. With the help
of these masses one can evaluate $V_4$ and $V_z$ in\linebreak
eq. (\ref{8}). Notice, that
working in dimensional regularization the leading order $\epsilon$-dependence
has to be kept in the plasma masses, because it gives finite
contributions to the order $g^4,\la^2$ due to one-loop divergences.
The scalar integral for $V_\ominus$ can be found in \cite{P}. More complicated
diagrams of this type can be reduced to the scalar case as described in
\cite{AE}. These calculations have to be extended to include all contributions
of order $g^4,\la^2$. After a long but straightforward calculation the
explicit formula for the potential in $\overline{\mbox{MS}}$-scheme is
obtained. This final result is given in the appendix.
Dropping the appropriate terms of $V$ the lower order
$g^4,\la$-result, as it is given by Arnold and Espinosa in \cite{AE}, can be
derived. We have found some minor discrepancies. A careful check of the
differences has shown that some misprints\footnote
   {$\sigma$ in eq.(8.2,8.3), eq.(A11) line 9,
    eq.(A19l), eq.(A19m) line 5,
    eq.(A19o) line 2, eq.(A25) line 2, eq.(A45)}
in \cite{AE} have to
be corrected to obtain complete agreement.

We have checked our full $g^4,\la^2$-result using the method of \cite{AE}.
Zero Matsubara frequency modes have been resummed and  the
necessary temperature counterterms have been calculated.
The obtained potential is in complete
agreement with the one we give in the appendix.

Note, that there are linear $\vi$-terms of fourth order in the couplings
present in $V_a$ and $V_z$. These terms cancel each other, thus ensuring the
relation $\lim_{\vi\to 0}\partial V/\partial\vi=0$ for all allowed
temperatures. This cancellation is essentially the same effect which leads to a
vanishing third order transverse gauge boson mass in the symmetric phase
\cite{BFHW}, as it can be seen in the contributions of diagrams fig. 6.o and
fig. 6.t of \cite{BFHW}.

The result of the present paper with the wave function correction
term of \cite{BBFH} gives the finite temperature effective action up to order
$g^4,\la^2$.

\subsection{Renormalization in the standard model}
In case of the standard model it is not possible to avoid the zero
temperature renormalization just by setting $\bar{\mu}=1/\beta$. The reason for
that is the large negative $g_Y^4\vi^4$-term, which dominates over the tree
level quartic term. This leads to
an $\overline{\mbox{MS}}$-potential unbounded from below for moderately large
top mass and small Higgs mass.

We perform a zero temperature renormalization in the on-shell scheme,
as described in \cite{BSH}. Higgs mass, top quark mass, W- and Z-boson
masses and the fine structure constant $\alpha$ are chosen as
physical parameters \cite{PDG}. The physical masses are the poles of the
propagators and $\alpha$ is defined in the Thompson limit. A multiplicative
renormalization of the coupling constants, the tree level Higgs mass square
$-\nu$ and the physical Higgs field is performed.
The wave function renormalization of the Higgs field is defined as usual by
    \be \delta Z_\vi=\frac{\partial}{\partial q^2}\mbox{Re}\;\Pi_\vi(q^2)
    \Big|_{q^2=m_{\vi,phys}^2}\, .\e
No wave function renormalization is needed for the other fields, because they
do not appear in the effective potential.
$v$ is defined to be the true vacuum
expectation value of the physical Higgs field. Therefore it needs no
corrections and no tadpole diagrams have to be considered.

The correction to the electric charge $\delta e$ is gauge independent
\cite{S}, as it can be easily checked explicitly using the results of
\cite{DS}. Therefore in the present calculation the formula for $\delta e$ from
\cite{BSH} is used. The logarithmic terms with the five light quark masses
are treated in the way described in \cite{Ho}, with data from \cite{J},
resulting in the vacuum polarization contribution:
    \be \mbox{Re}\,\Pi^{\gamma(5)}_{\mbox{\footnotesize had}}(M_Z^2)=
    -0.0282\pm 0.0009\, .\e
The dependence of the one-loop self
energy corrections on the gauge parameters has
been calculated in \cite{DS} for gauge bosons. Therefore the corrections in
Landau gauge, needed here, can be taken from \cite{DS,MS}. The self energy
corrections for the physical Higgs boson and the top quark can be easily
calculated in Landau gauge. Using these quantities the complete zero
temperature renormalization of the potential can be done. The result is
thereby freed of any dependence on $\bar{\mu}$.

Clearly, the analytic expression of these corrections to the potential is too
long to be given here. However, it seems worthwhile to give the
numerically most important parts of the corrections, to enable a simplified
usage of
the analytic result in the appendix. As it has already been mentioned, the
main contributions come from the $g_Y^4$-corrections to parameters of order
$\la$ (see also \cite{AE}):
    \be \delta\la=\frac{3g_Y^4}{8\pi^2}\ln\frac{m_t}{\bar{\mu}}\quad,\quad
    \delta\nu=\frac{3g_Y^4v^2}{16\pi^2}.\label{4}\e
Introducing this corrections in all terms in the potential contributing to
order $\la$ and using standard model tree level relations to calculate the
couplings one  obtains a result which is ``partially renormalized at zero
temperature''.
The corresponding correction to the $\overline{\mbox{MS}}$-potential reads
    \be\delta V=\frac{\vi^2}{2}\left(-\delta\nu+\frac{1}{2\beta^2}\delta\la
    \right)+\frac{\delta\la}{4}\vi^4\, .\label{10}\e
As we will see it later (sect.4), the numerical effect of this
simplification is not too severe.

\section{Results for pure SU(2)-Higgs model}
\subsection{Effective potential and surface tension}
To obtain an understanding of the qualitative effects of higher order
corrections we study first the pure SU(2)-Higgs model. In this section the
additional U(1)-symmetry and the effect of the fermions are neglected.
A discussion of this
simplified version may also be useful in view of lattice investigations, which
will probably deal with the pure SU(2)-Higgs model in the near future.

\setcounter{figure}{2}
\refstepcounter{figure}
\label{pot2}

\refstepcounter{figure}
\label{ofs1}

\refstepcounter{figure}
\label{ofs3}

\refstepcounter{figure}
\label{lh1}

\refstepcounter{figure}
\label{phi1}

\refstepcounter{figure}
\label{ofs6}

The relevant potential can be
easily derived from the formulas given in the appendix by performing the limit
$g_1,g_Y\longrightarrow 0$ and setting the number of families $n_f$ to zero.
Throughout this section standard model values for W-mass and vacuum expectation
value $v$ are used, unless stated otherwise : $m_W=80.22$ GeV and $v=251.78$
GeV.
The parameter $\bar{\mu}$ of dimensional regularization is set to $T=1/\beta$.
This can be justified by the small dependence on the renormalization procedure.
The differences between the results obtained in this scheme and in a scheme
with on-shell $T=0$ renormalization are very small. This phenomenon has been
observed in the Abelian Higgs model as well \cite{He}.

In fig. \ref{pot2} different approximations of the effective potential at their
respective critical temperatures are shown. The potentials to order
$g^3,\la^{3/2}$ and $g^4,\la$ can be obtained from \cite{Ca,BFHW} and \cite{AE}
respectively. Each approximation suggests a first order phase transition.
On the one hand the critical temperature and the position of the degenerate
minimum seems to be quite stable, on the other hand the hight of the barrier is
$\sim 10$ times larger for the $g^4,\la^2$ case than for the
$g^3,\la^{3/2}$-potential. No convergence of the perturbation
series can be claimed for the given parameters.

A more detailed picture
can be obtained by considering the surface tension \cite{CL}
\be \sigma=\int_0^{\vi_+}d\vi\sqrt{2V(\vi,T_c)}, \e
which may be seen as a measure of the strength of the phase transition. It can
be used conveniently to discuss the properties of the potential as a function
of the Higgs mass. The results are shown in fig. \ref{ofs1}.
For very small Higgs masses the third order potential gives a much
larger value for the surface tension than the more complete calculations.
The reason for that is the $g^4\vi^4$ contribution, which takes over the role
of the tree level $\la\vi^4$ term for small scalar coupling. This radiatively
induced quartic term ensures that $\sigma$ does not increase for small Higgs
masses, a maximum is found. As could have
been expected, corrections of higher order in $\la$ do not change the $g^4,\la$
result if the scalar mass is small.

This picture changes
drastically if larger Higgs masses are considered. In this region higher
order corrections produce an enormous increase in the surface tension. The
difference between the $g^4,\la^2$ and the $g^4,\la$ results looks very much
the same as in the case of the Abelian Higgs model (see the discussion in
\cite{He}). However, in contrast to the situation there in the SU(2)-model
both curves suggest
much larger values of the surface tension than the $g^3,\la^{3/2}$ result.

Let us compare first the results of order $g^3,\lambda^{3/2}$ and $g^4,\la^2$.
The increase in the strength of the phase transition,
studied already in \cite{BD}, can be traced back
to the infrared features of a non-Abelian gauge theory. The crucial
contribution is the one coming from the non-Abelian setting sun
diagram (fig. 2.m). It produces contributions to the potential of type
$\varphi^2\ln(\beta m_W)$ with negative sign. The huge effect of the
$\ln\beta m_W$-contribution to the coefficient of $g^4\vi^2$ can be understood
by recalling that at the critical temperature the leading order $\vi^2$-terms
essentially cancel. However, the $\vi^2\ln\vi$-type behaviour can not be
absorbed in a correction of $T_c$, these terms increase the strength of the
phase transition.
The effect becomes clear if one deletes the $\vi^2\ln\beta m_W$-term of
$V_m$ by hand.
The corresponding surface tension is shown in fig. 5 (long-dashed line).

We compare now the $g^4,\la$ \cite{AE} and the $g^4,\la^2$ results. The
complete calculation produces contributions of type
$\vi^2\ln\beta(m_W+m_{1,2})$ with positive sign (see appendix). These terms are
coming from the scalar-vector setting sun diagrams (fig. 2.a,b). In \cite{AE}
scalar masses have been neglected, resulting in spurious
$\vi^2\ln\beta m_W$-terms with positive sign, which reduces the  surface
tension. Another important contribution is the one proportional to
$g^2(m_1+3m_2)m_{WL}$ from $V_z$. This term comes
from scalar-vector diagrams of type of fig. 1.b and it was neglected in
\cite{AE}. On the relevant
scale ($\varphi<T$) it produces a very steep behaviour of the
potential, again increasing the surface tension.
The observed difference between the result of \cite{AE} and the
complete $g^4,\la^2$ calculation presented here is mostly due to these two
effects, together with the well known influence of the cubic scalar mass
contributions from $V_3$.

Another interesting effect of higher order $\la$-corrections is the complete
breakdown of the phase transition at a Higgs mass of about 100 GeV, where the
surface tension is very large. In this region the above mentioned
term, proportional to
$g^2(m_1+3m_2)m_{WL}$, becomes important. For a
temperature close to the uncorrected barrier temperature $T_b$, at which the
scalar masses vanish for $\vi=0$, it produces
an almost linear behaviour in the small $\vi$ region. This results in a
potential for which at $T=T_b$ the asymmetric minimum is not a
global minimum but only a local one. Note that $T_b$ is the lowest temperature
accessible in this calculation. In other words, the temperature region in which
the phase transition occurs can not be described by the given method, due to
infrared problems.

In order to illustrate the possible effects of the unknown infrared behaviour
of the transverse vector propagator, the dependence of the surface tension
on the magnetic mass can be studied. We follow the approach of \cite{BFHW},
where a magnetic mass motivated by the solution of the gap equations was
introduced. The transverse vector mass takes the form
    \be m_W^2=\left(\frac{g\vi}{2}\right)^2+\left(\frac{\gamma g^2}{3\pi\beta}
    \right)^2\, , \e
where $\gamma$ is some unknown parameter. One can introduce this redefined
transverse
mass in the most influential infrared contributions, i.e. in the
$m_W^3$- and in the $\vi^2\ln\beta m_W$-terms. We show in fig. \ref{ofs3} the
results obtained for $\gamma=$ 0, 2 and 4. The qualitative behaviour is
similar to results found in \cite{BFHW}. The main difference
is due to the fact
that the higher order result suggests a stronger first order phase
transition, thus for a given $m_{\mbox{\scriptsize Higgs}}$ a larger magnetic
mass is necessary to change the phase transition to second order.

A complete fourth order calculation of the surface tension has to include the
wave function correction term $Z_\vi(\vi^2,T)$ calculated in \cite{BBFH}. Using
the results of \cite{BBFH} we
have determined $\sigma$ for Higgs masses between 25-95 GeV. The numerical
effect of this $Z$-factor is very small, only $1\%-4\%$.

\subsection{Further properties of the potential}
The latent heat of the phase transition is another interesting quantity to be
calculated from the effective potential :
    \be \Delta Q=T\frac{\partial}{\partial T}V(\vi_+,T)\Big|_{T_c}\, ,\e
where $\vi_+$ is the position of the asymmetric minimum of $V$. We have plotted
$\Delta Q$ as a function of $m_{\mbox{\scriptsize Higgs}}$ in fig. \ref{lh1}.
The latent heat of the higher order calculations ($g^4,\la$ and $g^4,\la^2$)
increases almost linearly with the Higgs mass.
This somewhat surprising behaviour can be understood by observing that for
those potentials neither the position of the degenerate minimum nor the height
of the barrier change significantly with increasing Higgs mass
(see fig. \ref{ofs1}). On the other hand the critical temperature is
essentially proportional to $m_{\mbox{\scriptsize Higgs}}$.

For completeness, the quantity $\vi_+/T_c$, relevant for baryogenesis, is shown
in fig. \ref{phi1} as a function of the Higgs mass. It is interesting to
observe that the upper part of the region favouring baryogenesis \cite{CKN},
i.e. $\vi_+/T_c
\approx 1$ at $m_{\mbox{\scriptsize Higgs}}\approx 40$ GeV, coincides with the
region of best reliability of the perturbative approach. As has already been
pointed out in \cite{AY,He}, this parameter does not reflect the dramatic
change of the potential at critical temperature introduced by higher order
corrections.

Now the question arises whether a good convergence of the perturbation series,
which can not be claimed in the whole range of $\la$ for a realistic gauge
coupling $g=0.64$, could be present in the region of much smaller gauge
coupling constants. This seems indeed to be the case, as can be seen in fig.
\ref{ofs6}, where the surface tensions of order $g^3,\lambda^{3/2}$ and
$g^4,\lambda^2$
are plotted for a model with a vector mass of 20 GeV, i.e. $g=0.16$. In the
used Higgs mass range the two results for $\sigma$ differ by a factor of two at
most. The relative size of this range,
i.e. the ratio of the minimal and maximal values of the Higgs mass,
is 4, which is twice as large as the  range for
the model with  $m_W=80$ GeV.

\refstepcounter{figure}
\label{ofs4}

\refstepcounter{figure}
\label{phi2}

\section{Standard model results}
In the case of the full standard model the
qualitative behaviour of the potential is essentially the same as for the
SU(2)-Higgs model. The main difference is a decrease of the surface tension
by a factor $\sim 4$. This can be traced back to the large top mass. Also the
characteristic points of the surface tension plot of fig. \ref{ofs1} are
shifted to higher values of the Higgs mass. We show $\sigma$ as a function of
$m_{\mbox{\scriptsize Higgs}}$ in fig. \ref{ofs4}. The complete breakdown of
the $g^4,\la^2$ calculation, observed at
$m_{\mbox{\scriptsize Higgs}}\approx 100$ GeV for the pure SU(2) case, occurs
at $m_{\mbox{\scriptsize Higgs}}\approx 200$ GeV in the full model.
These quantitative differences do not change the qualitative features
of the potential, thus
the discussion given in the previous section
does also apply to the standard model. The difference between the fully
renormalized potential and the partially renormalized potential
(see eq. (\ref{4}),(\ref{10})) is not too severe in view of the huge
uncertainties still present in the
perturbative approach. Again, the position of the
second minimum at the critical temperature, given in fig. \ref{phi2}, does not
depend as strongly on the order of the calculation as the height of the
barrier. Unfortunately, the region \linebreak
$m_{\mbox{\scriptsize Higgs}}\approx 40$ GeV, in which the
reliability of the
perturbative approach is the best and
$\vi_+/T_c\approx 1$, is well below the
experimental Higgs mass bound.

\section{Conclusions}
In the previous sections we have calculated and analyzed the finite temperature
effective potential of the standard model up to order $g^4,\la^2$. We have
determined several physical quantities as functions of the Higgs mass.
However, to the given order
the systematic expansion in coupling constants does not permit a definitive
statement about the character of the phase transition for realistic Higgs
masses. This is seen from the fact that the $g^4,\la^2$-corrections are
huge and even the step from a $g^4,\la$-calculation to the complete
$g^4,\la^2$-result changes the potential essentially if the Higgs mass is
large. One source of the dramatic increase of the surface tension are the
infrared contributions of the typical non-Abelian diagrams.
Quantitative information on a possible infrared cutoff, e.g. a magnetic
mass term, could increase the reliability of the calculation drastically.
For large Higgs masses another infrared problem is connected with the scalar
sector, namely the corrected leading order scalar masses vanish near the
critical temperature producing an almost linear term in the potential.
It has to be concluded, that although resummation techniques permit
a systematic expansion
in coupling constants, the numerical results still point to the unknown
low momentum behaviour of the theory as the main obstacle of any reliable
prediction.\\[1cm]
Special thanks go to W. Buchm\"uller for his continuous
support of this work.
Helpful discussions with D. B\"odeker, T. Helbig, B. Kniehl and H. Kohrs
are also acknowledged. Z. F. was partially supported by Hung. Sci. Grant under
Contract No. OTKA-F1041/3-2190.

\section*{Appendix}
Here the different contributions to the potential described in sect. 2 are
given explicitly. The formulas have been simplified as much as possible to
enable a direct numerical use and further analytic investigation.
The authors\footnote{e-mail fodor@vxdesy.desy.de or t00heb@dhhdesy3.bitnet}
are ready to supply a FORTRAN code evaluating the different approximations
of the effective potential ($g^3,\la^{3/2}$; $g^4,\la$; $g^4,\la^2$
pure $SU(2)$ and standard model, with complete on-shell renormalization
or partial renormalization)
as a function of $\vi$ and $T$.

Linear mass terms, poles in $2\epsilon=4-n$ and
terms proportional to the constant $\iota_\epsilon$ (see \cite{AE}), which
cancel systematically in the final result, are not shown and the limit
$n\rightarrow 4$ has already been performed. The leading order
resummed scalar masses are given by
    \be m_1^2=2\la\vi^2+m_2^2\quad,\quad m_2^2=\la\vi^2-\nu+\frac{1}{12\beta^2}
    \left(6\la+\frac{9}{4}g_2^2+\frac{3}{4}g_1^2+3g_Y^2\right)\, ,\e
while the transverse vector boson masses and the fermion mass remain
uncorrected to leading order :
    \be m_W=\frac{1}{2}g_2\vi\quad,\quad m_Z=m_W/\cos \theta_W\quad,
    \quad m_f=\frac{1}{\sqrt{2}}g_Y\vi\, .\e
The longitudinal SU(2)$\times$U(1) mass matrix receives temperature
corrections in the diagonal elements \cite{Ca}
    \be m_{WL}^2=\frac{1}{4}g_2^2\vi^2+\frac{g_2^2}{\beta^2}\left(\frac{5}{6}
    +\frac{1}{3}n_f\right)\quad,\quad m_{BL}^2=\frac{1}{4}g_1^2\vi^2+
    \frac{g_1^2}{\beta^2}\left(\frac{1}{6}+\frac{5}{9}n_f\right)\, ,\e
which result in longitudinal masses defined by
    \be \left(\begin{array}{cc}
        m_{WL}^2 & -\mif{1}{4}g_1g_2\vi^2  \\&\\
       -\mif{1}{4}g_1g_2\vi^2          & m_{BL}^2
    \end{array}\right)       =
    \left(\begin{array}{cc}
        \cos\tilde{\theta} & \sin\tilde{\theta} \\&\\
       -\sin\tilde{\theta} & \cos\tilde{\theta} \end{array}\right)
    \left(\begin{array}{cc}
        m_{ZL}^2 & 0 \\&\\ 0 & m_{\gamma L}^2 \end{array}\right)
    \left(\begin{array}{cc}
        \cos\tilde{\theta} & -\sin\tilde{\theta} \\&\\
        \sin\tilde{\theta} & \cos\tilde{\theta} \end{array}\right).\e
In the following the short hand notations
    \be s=\sin \theta_W\quad,\quad c=\cos \theta_W\quad,\quad \tilde{s}=\sin
    \tilde{\theta}\quad,\quad \tilde{c}=\cos \tilde{\theta} \e
are used. Evaluating the scalar one- and two-loop temperature integrals the
constants
    \be c_0=\mif{3}{2}+2\ln{4\pi}-2\gamma\approx 5.4076\quad\mbox{and}\quad
    c_2\approx 3.3025\e
are introduced following \cite{DJ} and \cite{P} respectively. Now all the
contributions to the potential, which have to be
summed according to formulas (\ref{8}) and (\ref{9}), can be given
explicitly :

\begin{eqnarray}
  V_3&\!\!\!=&\!\!\!\frac{\vi^2}{2}\left[-\nu+\frac{1}{\beta^2}\left(\frac{1}
  {2}\la+\frac{3}{16}g_2^2+\frac{1}{16}g_1^2+\frac{1}{4}g_Y^2\right)\right]
  +\frac{\la}{4}\vi^4\\
&&\nonumber\\
  &&\!\!\!\!\!\!\!\!\!\!\!\!-\frac{1}{12\pi\beta}\left[{ m_1}^{3}+3\,
  { m_2}^{3}+4\,{ m_W}^{3}+2\,{ m_{WL}}^{3}+2\,{ m_Z}^{3}+{ m_{ZL}}^{3}+
  { m_{\gamma L}}^{3}\right]\nonumber
\\
&&\nonumber\\
&&\nonumber\\
  V_a&\!\!\!=&\!\!\!\frac{g_2^2}{32\pi^2\beta^2}\;\Bigg[m_W^2\left(2-\frac{1}
  {c^2}+\frac{1}{2c^4}\right)\left(\ln{\frac{\beta}{3}}-\frac{1}{12}\ln{
  \bar{\mu}^2\beta^2}-\frac{1}{6}c_0+\frac{1}{4}c_2+\frac{1}{4}\right)\\
&&\nonumber\\
  &&\!\!\!\!\!\!\!\!\!\!\!\!+\frac{1}{4}\left(1+\frac{1}{2c^2}\right)\left\{
  \left(m_1^2+3m_2^2\right)\left(-4\ln{\frac{\beta}{3}}+\ln{\bar{\mu}^2\beta^2}
  -c_2\right)+2m_2\left(m_1+m_2\right)\right\}\nonumber\\
&&\nonumber\\
  &&\!\!\!\!\!\!\!\!\!\!\!\!-4s^2m_2^2\ln{(2m_2)}-\frac{1}{2m_W}\left(1+\frac
  {1}{2c}\right)(m_1-m_2)^2(m_1+m_2)+2m_2m_Zs^2\nonumber\\
&&\nonumber\\
  &&\!\!\!\!\!\!\!\!\!\!\!\!-\frac{m_W}{2}\left(1+\frac{1}{2c^3}\right)
  (m_1+3m_2)-\frac{3}{4m_W^2}\left(m_1^2-m_2^2\right)^2\ln(m_1+m_2)\nonumber\\
&&\nonumber\\
  &&\!\!\!\!\!\!\!\!\!\!\!\!+\frac{1}{2m_W^2}\left\{m_W^4-2\left(m_1^2+m_2^2
  \right)m_W^2+\left(m_1^2-m_2^2\right)^2\right\}\ln{(m_1+m_2+m_W)}\nonumber\\
&&\nonumber\\
  &&\!\!\!\!\!\!\!\!\!\!\!\!+\frac{1}{4m_W^2}\left\{m_Z^4-2\left(m_1^2+m_2^2
  \right)m_Z^2+\left(m_1^2-m_2^2\right)^2\right\}\ln{(m_1+m_2+m_Z)}\nonumber\\
&&\nonumber\\
  &&\!\!\!\!\!\!\!\!\!\!\!\!+\left(\frac {1}{4c^2}-s^2\right)\left(m_Z^2-4
  m_2^2\right)\ln(2m_2+m_Z)+\frac{1}{2}\left(m_W^2-4m_2^2\right)\ln(2m_2+m_W)
  \Bigg]\nonumber
\\
&&\nonumber\\
&&\nonumber\\
V_b&\!\!\!=&\!\!\!\frac{g_2^2}{64\pi^2\beta^2}\Bigg[\left\{\left(c^4+\frac{1}
  {c^4}+4c^2+\frac{4}{c^2}-10\right)m_W^2+2\left(c^2-\frac{1}{c^2}\right)s^2
  m_2^2+\frac{s^4m_2^4}{m_W^2}\right\}\\
&&\nonumber\\
  &&\!\!\!\!\!\!\!\!\!\!\!\!\times\ln(m_W+m_Z+m_2)+\left\{\left(5-4c^2\right)
  m_W^2-\frac{1}{m_W^2}\left(m_W^2c^2+m_2^2s^2\right)^2\right\}\ln(m_W+m_2)
  \nonumber\\
&&\nonumber\\
  &&\!\!\!\!\!\!\!\!\!\!\!\!-\frac{s^4}{m_W^2}\left(m_Z^2-m_2^2\right)^2\ln(
  m_Z+m_2)+m_W\left\{m_2\left(\frac{1}{c^2}-c^2+\frac{s^4}{c}
  \right)+m_1\left(1+\frac{1}{2c^3}\right)\right\}\nonumber\\
&&\nonumber\\
  &&\!\!\!\!\!\!\!\!\!\!\!\!+m_W^2\left\{\left(\frac {5}{2c^2}+\frac{5}{8c^4}
  -\frac{5}{4}\right)\left(2\ln\frac{\beta}{9\bar{\mu}}+c_2\right)+
  \frac{c^2}{2}-\frac{5}{2}+\frac{2}{c^2}+\frac{s^2}{c}\left(c^2-
  \frac{1}{c^2}\right)\right\}\nonumber\\
&&\nonumber\\
  &&\!\!\!\!\!\!\!\!\!\!\!\!-\frac{1}{2m_W^2}\left\{2\left(m_W^2-m_1^2\right)^2
  \ln(m_W+m_1)+\left(m_Z^2-m_1^2\right)^2\ln(m_Z+m_1)\right\}\nonumber\\
&&\nonumber\\
  &&\!\!\!\!\!\!\!\!\!\!\!\!+\!\!\left(4m_W^2-2m_1^2+\frac
  {m_1^4}{2m_W^2}\right)\ln(2m_W+m_1)+\frac{1}{c^2}\left(2m_Z^2
  -m_1^2+\frac{m_1^4}{4m_Z^2}\right)\ln(2m_Z+m_1)\nonumber\\
&&\nonumber\\
  &&\!\!\!\!\!\!\!\!\!\!\!\!+s^2m_2^2\left(2\ln m_2+
  \frac{s^2}{c}\right)+\frac{m_1^2}{2}\left(1+\frac{1}{2c^2}\right)+\frac{1}
  {m_W^2}\left(s^4m_2^4\ln m_2+\frac{3}{4}m_1^4\ln m_1\right)\nonumber\\
&&\nonumber\\
  &&\!\!\!\!\!\!\!\!\!\!\!\!
  +\frac{\vi^2}{4g_2^2}\Big\{(g_2\tilde{c}+g_1\tilde{s})^4\ln(2m_{ZL}+m_1)
  +(g_2\tilde{s}-g_1\tilde{c})^4\ln(2m_{\gamma L}+m_1)\nonumber\\
&&\nonumber\\
  &&\!\!\!\!\!\!\!\!\!\!\!\!+2g_2^4\ln(2m_{WL}+m_1)
  +2(g_2\tilde{c}+g_1\tilde{s})^2(g_2\tilde{s}-g_1\tilde{c})^2\ln(m_{ZL}+m_{
  \gamma L}+m_1)\nonumber\\
&&\nonumber\\
  &&\!\!\!\!\!\!\!\!\!\!\!\!
  +4g_2^2g_1^2\left(\tilde{s}^2\ln(m_{WL}+m_{ZL}+m_2)
  +\tilde{c}^2\ln(m_{WL}+m_{\gamma L}+m_2)\right)\Big\}\Bigg]\nonumber
\\
&&\nonumber\\
&&\nonumber\\
  V_i&\!\!\!=&\!\!\!\frac{1}{8\pi^2\beta^2}\Bigg[\left\{\frac{g_2^2m_f^2}{96}
  \left(10+\frac{17}{c^2}\right)+\frac{g_2^2n_fm_W^2}{36}\left(\frac{10}{c^2}
  -\frac{5}{c^4}-14\right)+g_s^2m_f^2\right\}\\
&&\nonumber\\
  &&\!\!\!\!\!\!\!\!\!\!\!\!\times(\ln{\bar{\mu}^2\beta^2}-c_0+\frac{1}{2}
  +\frac{10}{3}\ln2)-\frac{4g_2^2n_fm_W^2}{27}\left(\frac{10}{c^2}-\frac{5}
  {c^4}-14\right)\ln2\Bigg]\nonumber
\\
&&\nonumber\\
&&\nonumber\\
  V_j&\!\!\!=&\!\!\!\frac{g_Y^2}{128\pi^2\beta^2}\left[\left(9m_f^2-m_1^2-
3m_2^2\right)\left(\ln{\bar{\mu}^2\beta^2}-c_0+\frac{3}{2}-\ln4\right)+48m_f^2
  \ln2\right]
\\
&&\nonumber\\
&&\nonumber\\
  V_m&\!\!\!=&\!\!\!\frac{g_2^2}{16\pi^2\beta^2}\Bigg[m_W^2\Bigg(\Big(-\frac
  {1}{4c^4}-\frac{1}{c^2}+\frac{5}{2}-c^2-\frac{c^4}{4}\Big)\ln(m_W+m_Z)\\
&&\nonumber\\
  &&\!\!\!\!\!\!\!\!\!\!\!\!+\left(\frac{1}{8c^4}+\frac{1}{c^2}-5-4c^2\right)
  \ln(2m_W+m_Z)+\frac{31}{8}\ln{\bar{\mu}^2\beta^2}-\frac{11}{16}c_0-\frac
  {51}{16}c_2-\frac{251}{96}\nonumber\\
&&\nonumber\\
  &&\!\!\!\!\!\!\!\!\!\!\!\!-\frac{11}{12}c-\frac{5}{4c}+\frac{1}{8c^2}-4s^2
  \ln2+\frac{1}{4}c^2(c-\frac{1}{2})+\frac{1}{8}\left(2-\frac{1}{c^4}\right)
  \ln c-\frac{51}{4}\ln\frac{\beta}{3}\nonumber\\
&&\nonumber\\
  &&\!\!\!\!\!\!\!\!\!\!\!\!+\left(\frac{1}{8c^4}-\frac{23}{4}+5c^2+\frac{1}
  {4}c^4\right)\ln m_W\Bigg)-m_Wm_{WL}(1+c)-2s^2m_{WL}^2\ln(2m_{WL})
  \nonumber\\
&&\nonumber\\
  &&\!\!\!\!\!\!\!\!\!\!\!\!+\left(\frac{1}{2}m_W^2-2c^2m_{WL}^2\right)\ln(2
  m_{WL}+m_Z)+\frac{1}{2}m_{WL}^2-\frac{m_{WL}^3}{m_W}\nonumber\\
&&\nonumber\\
  &&\!\!\!\!\!\!\!\!\!\!\!\!+\tilde{s}^2\Bigg\{\left(m_W^2-2m_{WL}^2-
  2m_{\gamma L}^2+\frac{\left(m_{WL}^2-m_{\gamma L}^2\right)^2}{m_W^2}\right)
  \ln(m_{WL}+m_{\gamma L}+m_W)\nonumber\\
&&\nonumber\\
  &&\!\!\!\!\!\!\!\!\!\!\!\!+m_{\gamma L}(m_{WL}-m_W)+\frac{
  m_{WL}^2m_{\gamma L}+m_{WL}m_{\gamma L}^2-m_{\gamma L}^3}{m_W}
  -\frac{\left(m_{WL}^2-m_{\gamma L}^2\right)^2\ln(m_{\gamma L}
  +m_{WL})}{m_W^2}\Bigg\}\nonumber\\
&&\nonumber\\
  &&\!\!\!\!\!\!\!\!\!\!\!\!+\tilde{c}^2\Bigg\{\hspace{1cm}
  m_{\gamma L}\longrightarrow m_{ZL} \hspace{1cm}\Bigg\}\Bigg]\nonumber
\\
&&\nonumber\\
&&\nonumber\\
  V_p&\!\!\!=&\!\!\!-\frac{3\la^2\vi^2}{32\pi^2\beta^2}\left[\ln
  \frac{9\bar{\mu}^2}{\beta^2}-c_2-2\ln\{m_1(m_1+2m_2)\}\right]
\\
&&\nonumber\\
&&\nonumber\\
  V_4&\!\!\!=&\!\!\!\frac{\vi^2}{64\pi^2\beta^2}\Bigg[\frac{g_2^4}{4}\left\{
  \left(\frac{1}{c^2}-\frac{1}{2c^4}-\frac{35}{4}\right)(
  \ln{\bar{\mu}^2\beta^2}-c_0)+\frac{293}{72}-\frac{1}{18c^2}
  -\frac{13}{18c^4}\right\}\\
&&\nonumber\\
  &&\!\!\!\!\!\!\!\!\!\!\!\!+\frac{g_2^4n_f}{27}\left(\frac{10}{c^2}
  -\frac{5}{c^4}-14\right)+g_Y^2\left(\frac{g_2^2s^2}{3c^2}-\frac{3g_Y^2}{4}
  +4g_s^2\right)\ln4+g_2^2\left(\frac{g_Y^2}{2}+\la\right)\left(1+\frac{1}
  {2c^2}\right)\nonumber\\
&&\nonumber\\
  &&\!\!\!\!\!\!\!\!\!\!\!\!+3g_Y^2\la\left(\ln{\bar{\mu}^2\beta^2}-c_0+\frac
  {3}{2}\right)\Bigg]-\frac{1}{64\pi^2}\Big\{-4m_W^4-2m_Z^4-24m_f^4\ln4
  \nonumber\\
&&\nonumber\\
  &&\!\!\!\!\!\!\!\!\!\!\!\!+\left(6m_W^4
  +3m_Z^4+m_1^4+3m_2^4-12m_f^4\right)\left(
  \ln{\bar{\mu}^2\beta^2}-c_0+\frac{3}{2}\right)\Big\}\nonumber
\\
&&\nonumber\\
&&\nonumber\\
  V_z&\!\!\!=&\!\!\!\frac{1}{32\pi^2\beta^2}\Bigg[\mif{1}{4}(m_1+3m_2)\Big\{
  g_2^2\left(2m_{WL}+4m_W+m_{ZL}\tilde{c}^2+2m_Zc^2\right)\\
&&\nonumber\\
  &&\!\!\!\!\!\!\!\!\!\!\!\!+g_1^2\left(m_{ZL}\tilde{s}^2+2m_Zs^2\right)
  +m_{\gamma L}\left(g_2^2\tilde{s}^2+g_1^2\tilde{c}^2\right)\Big\}\nonumber\\
&&\nonumber\\
  &&\!\!\!\!\!\!\!\!\!\!\!\!+\mif{1}{2}g_2g_1(m_1-m_2)\left\{2m_Zsc+(m_{ZL}
  -m_{\gamma L})\tilde{s}\tilde{c}\right\}+4g_2^2m_{WL}\left(m_W
  +m_Zc^2\right)\nonumber\\
&&\nonumber\\
  &&\!\!\!\!\!\!\!\!\!\!\!\!+g_2^2m_W\left(4m_{\gamma L}\tilde{s}^2+
  \mif{8}{3}m_W+\mif{16}{3}m_Zc^2+4m_{ZL}\tilde{c}^2\right)+3\la\left(
  m_1m_2+\mif{1}{2}m_1^2+\mif{5}{2}m_2^2\right)\Bigg].\nonumber
\end{eqnarray}

\end{document}